# GLOBULAR CLUSTERS IN THE INNER REGIONS OF NGC 5128 (CEN A)[1]

Dante Minniti

*Lawrence Livermore National Laboratory, MS L-413, P.O.Box 808, Livermore, CA 94550,*
*and European Southern Observatory, D–85748 Garching bei München, Germany*

M. Victoria Alonso

*Observatorio Astronomico Cordoba, Laprida 854, 5000 Cordoba, Argentina,*
*and European Southern Observatory, D–85748 Garching bei München, Germany*

Paul Goudfrooij

*European Southern Observatory, D–85748 Garching bei München, Germany*

Pascale Jablonka

*Observatoire de Paris, Section de Meudon, F–92195 Meudon Cedex, France*

Georges Meylan

*European Southern Observatory, D–85748 Garching bei München, Germany*

---

## ABSTRACT

We have identified 26 new globular cluster candidates in the inner 3 kpc of NGC 5128 (Cen A), the nearest known large galaxy being the probable product of a merger. The clusters are selected on the basis of their structural parameters (observed core diameters and ellipticities), as measured from archival WFPC HST images. IR photometry obtained with IRAC2B at the ESO/MPI 2.2–m telescope is combined with the optical HST photometry.

Most of these clusters have normal colors typical of old globular clusters like those found in the Milky Way and M31. We estimate their metal abundances based on the $R - K_0$ color, confirming the existence of a metallicity gradient in the inner regions of NGC 5128. The presence of metal–rich globular clusters suggests that one of the colliding galaxies was a bulge–dominated galaxy (E or early S).

A few clusters have colors and magnitudes similar to intermediate–age clusters containing carbon stars in the Magellanic Clouds. If the intermediate–age clusters were formed during a merger, then this episode must have occurred a few Gyr ago. Alternatively, we are looking at the cluster members of one of the colliding galaxies, which would then have been a late–type disk galaxy.

*Subject headings:* Clusters: globular — Galaxies: individual — Galaxies: abundances — Galaxies: formation

## 1. Introduction

The study of globular cluster systems in galaxies is crucial in our understanding of galaxy formation and evolution. Schweizer (1986) proposed that globular clusters may be formed during galaxy merger episodes. In distant galaxies, the observational evidence in favor of globular cluster formation during merger episodes is rapidly growing (*e.g.* Holtzman *et al.* 1992, Zepf & Ashman 1993, Whitmore & Schweizer 1995). This mechanism naturally accounts for the high specific frequency of globular clusters observed in elliptical galaxies (e.g., Harris 1991, Ashman & Zepf 1992).

A very special case for the study of globular cluster formation in such violent events is NGC 5128 (Cen A), the nearest giant galaxy which is probably the product of a merger (e.g. Baade & Minkowski 1954, Sérsic 1982, Mathieu *et al.* 1995). This galaxy has a rich system of globular clusters (Hesser *et al.* 1984), with membership confirmed spectroscopically (Sharples 1988). Frogel (1984) obtained near–IR photometry for 12 of the brightest clusters in this galaxy ($14 \leq K \leq 15$), finding that some of them have $[Fe/H] \geq 0$. Harris *et al.* (1992) derive metallicities for 62 bright clusters based on Washington photometry. In particular, they do not confirm the high metallicities found by Frogel (1984), a conclusion that is supported by Jablonka *et al.* (1995), based on spectrophotometry of the 5 most metal–rich clusters in Frogel's sample. Another important characteristic of the globular cluster system of NGC 5128 is the broad color distribution found by Harris *et al.* (1992). Zepf & Ashman (1993) first suggested that this color distribution is in fact bimodal, which is probably due to a past merger event. Another plausible cause for broad color distributions of globular cluster systems is on-going cluster formation in cooling flows (*e.g.* Nørgaard–Nielsen et al. 1993), but NGC 5128 has no cooling flow (Forman et al. 1985). An independent confirmation of the bimodality is found by Hui et al. (1995), who showed that the metal–rich cluster system in NGC 5128 rotates faster than the metal–poor cluster system.

However, there are only very few known clusters in the inner regions of this galaxy (Sharples 1995), and all previous works acknowledge their incompleteness in these regions. Consequently, the mere existence of super–metal–rich clusters in NGC 5128 remains to be confirmed. These putative super–metal–rich clusters are expected to be found preferentially in the inner regions, since in our Galaxy about 90% of the globular clusters with $[Fe/H] \geq -0.8$ are confined to the inner 3 kpc (Minniti 1995). Indeed, Harris *et al.* (1992) concluded that the innermost clusters are redder and therefore more metal–rich, but stated that this result is marginal due to the complicated structure of the dusty inner regions of NGC 5128.

The usual problems inherent to the identification and study of globular clusters in the inner regions of galaxies beyond the Local Group are faintness, crowding, high background galaxy light, nonuniform extinction, foreground contamination by stars in our own galaxy, and background contamination by distant galaxies. In this paper we take advantage of the combination of archival HST data with our ground–based IR observations in order to overcome these difficulties. We report on the identification of 32 globular cluster candidates in the inner 3 kpc of the peculiar E2 galaxy NGC 5128, of which 26 clusters were hitherto unknown. This new sample of clusters, deep into the core of NGC 5128, represents a basis for the determination of the metallicity gradient in the globular cluster system in this galaxy, and for the study of the history of cluster formation in a probable recent merger event.

## 2. Identification of Globular Clusters from the WFPC Images

We use archival images obtained with the WFPC and the pre–COSTAR HST in order to identify globular clusters in the inner regions of NGC 5128. These images consist of a set of 1200 sec exposures taken through the F675W filter with different polarization angles, the combination of which yields an unpolarized light image of 6000 sec. The magnitudes for the sources were obtained by aperture photometry with 4 pixel radius (enclosing the central peak containing $\sim 20\%$ of the flux), and with the sky background defined in an annulus between 10 and 15 pixels. Typical aperture corrections (0.92 mag) were obtained using five known clusters in the field (G206, G268, G359, G169, and G292), instead of stellar sources for which the PSF is different, and applied to the whole sample.

We have calibrated the HST frames with existing ground based photometry (Harris et al. 1992) in the $T_1$ filter for a few known globular clusters in this region. Unfortunately, the cluster G242 falls right in between two CCD chips of the WFPC, and we cannot use it for the calibration. We are left with clusters G206, G268, and G395, for which our final transformation gives $R = T_1 = F675W + 0.134$, with $\sigma = 0.08$. This error is smaller than our estimated zero point error, which we take as the $rms$ difference with the magnitude of the cluster G395, $\sigma = 0.13$ mag.

Note that Harris et al. (1992) comment that cluster G395 is located next to a charge overflow from a brighter image, which makes its $T_1$ magnitude more uncertain. This is the single most important contribution to the error budget in the $R$ photometry, far larger than the errors in the transformations between the different filter systems, than the color terms, and than the internal $\sqrt{N}$ photometric errors.

We note that the F675W filter is essentially equivalent to the Cousins $R$ filter: Harris et al. (1993) derive the transformation $R = F675W - 0.163\,(V-I) + 0.0229\,(V-I)^2$ with $\sigma = 0.012$. The Cousins $R$ filter is, in turn, essentially equivalent to the Washington $T_1$ filter: Taylor (1986) gives $R - I = 0.948\,T_1 - T_2 + 0.03$ with $\sigma = 0.002$, and Geisler (1996) gives $R = T_1 + 0.003 - 0.017\,(C - T_1)$ with $\sigma = 0.02$ valid for the color range of interest here ($-1.3 \leq C - T_1 \leq 2.2$).

The typical core diameters of Galactic globular clusters are $D_c = 6$ pc, with a total spread from about 0.2 to 20 pc, as listed in the most recent compilation by Harris (1995). At a distance of 3 Mpc this diameter translates into $\sim 0.32$ arcsec (see compilation of distances in Table 1 of Shopbell et al. 1993). These typical sizes can be resolved thanks to the superb spatial resolution of the WFPC (0.1 $arcsec/pix$), giving $D_c \approx 3$ pixels. Thus, foreground stars can be distinguished from typical globular clusters in NGC 5128. Furthermore, Galactic globular clusters are almost round White & Shawl 1987), which gives an extra criterion for the discrimination against background galaxies. For example, the ellipticity of the most massive Galactic globular cluster, $\omega$ Cen, is $\epsilon = 0.12$ (Meylan & Mayor 1986). Also, the most extreme case of a flattened cluster in the Large Magellanic Cloud is NGC 1978, with ellipticity $\epsilon = 0.3$

(Fischer *et al.* 1992).

Thus, the basic procedure we use for identifying globular clusters consists of looking at their sizes and ellipticities. Figure 1 shows a plot of ellipticity $\epsilon$ *vs.* FWHM of sources in the PC4 frame, where three globular clusters which are confirmed members of NGC 5128 (based on their radial velocities) are identified. Note that this procedure clearly produces an incomplete sample of globular clusters, where extremely compact or very loose ones are missed. However, based on their morphology, our candidates can be considered *bonafide* globular clusters.

Using these morphological selection criteria, the present sample is not biased by color selection. Previous samples were biased in this respect, and therefore unable to reveal the existence of a metallicity gradient in the inner regions of this galaxy.

## 3. The IR Photometry

The HST images are complemented with our deep near–IR images, which have the advantage of high contrast and reduced reddening sensitivity. Our IR mosaics are located in the central region, within the effective radius of NGC 5128 ($r_e = 5\ kpc$, Dufour *et al.* 1979). The innermost structure of NGC 5128 is complicated by the presence of large and non–uniform extinction, and the surface brightness of the underlying galaxy ranges from $\mu_K = 12$ to 16 $mag\ arcsec^{-2}$ in the region observed here (Quillen *et al.* 1993). Therefore, it would have been challenging to try to identify the clusters on top of the galaxy light without the aid of the HST images.

Our new IR observations were obtained with IRAC2B at the ESO/MPI 2.2–m telescope on March 4, 1995. We mapped a region of $3 \times 3$ arcmin$^2$ with the $JHK'$ filters, centered at about 1 arcmin NE of the nucleus of the galaxy.

The photometric reductions were done using the aperture photometry packages in IRAF. The calibration is based on 3 standard stars of Elias *et al.* (1988) observed during the night. These calibrations give errors $\sigma_J = 0.03$, $\sigma_H = 0.03$, and $\sigma_K = 0.05$. The achieved limiting magnitudes are $K = 18\ mag$ and $J = 19.5\ mag$, which are complementary to those of the WFPC images, $R = 21\ mag$. These limiting magnitudes reach clusters $\sim 1\ mag$ fainter than the peak of the globular cluster luminosity function in NGC 5128.

The positions of the globular cluster candidates are displayed in Figure 2 (Plate 1), along with the location of the HST frame. The cluster IDs, coordinates, radial distances from the nucleus, and observed magnitudes and colors are listed in Table 1. The $x$ and $y$ coordinates are measured in pixels ($scale = 0.49\ arcsec/pix$) with origin in the lower left corner of the mosaic, and the radial distances $r$ are measured in arcmin.

Reddening maps are constructed by dividing the different color frames by one another. Our reddening maps look similar to those of Meadows & Allen (1992) and of Quillen *et al.* (1993). Based on these maps, it was decided to study the candidates away from the inner region enclosed by the prominent dust lane, shown in Figure 2, where H II regions and young stellar associations are present (e.g. Dufour *et al.* 1979, Graham 1979, Nicholson *et al.* 1992). Even though the majority of the sources were detected in the dust lane region, it is not trivial to evaluate the amount of reddening that affects these different sources, and we will not discuss them here.

Following Frogel (1984), we adopt a to-

tal reddening of $E(B-V) = 0.11$, (equivalent to $A_K = 0.04$, $E(J-K) = 0.05$ and $E(R-K) = 0.22$) for the clusters observed here. By comparison with the color–color diagram of the background field, we estimate that only $E(J-K) \leq 0.02$ is due to the internal reddening of NGC 5128 itself.

We also mapped in $JHK'$ a comparison field located at about 30 arcmin NE of the nucleus, in order to estimate the percentage of back- and foreground contaminations. $JHK'$ photometry in this comparison field allows us to estimate the completeness of the present sample. We identified 32 globular clusters based on their morphology. Matching the number counts in the NGC 5128 field with those of the comparison field gives a total of 50 expected globular clusters in the region studied down to $K = 18$ $mag$. Based on the galaxy counts of Tyson (1988) and Hu & Ridgway (1994), there should be only about 5 galaxies present in our field, most of which would have been eliminated from our list based on their morphology (at most one of them would be an elliptical or compact galaxy – $cf$. Griffiths $et$ $al.$ 1994).

Figure 3 shows the optical–IR $K$ $vs.$ $R-K$ color–magnitude diagram for the cluster candidates. Their location in this diagram and in the color–color diagrams (Figures 4 and 5) matches the locus expected for globular clusters found in Local Group galaxies. Therefore, from their sizes and shapes as well as from their colors, the candidates in our list seem to be genuine globular clusters.

## 4. The Old Globular Clusters

Aaronson & Malkan (1978) showed that there is a clear correlation between the optical–IR colors and the metallicity for Galactic globular clusters. This correlation in the $V-K$ $vs.$ $[M/H]$ diagram is valid in the range $-2.2 \leq [M/H] \leq -0.3$. A more recent calibration, valid within $-2.5 \leq [M/H] \leq +0.5$ is given by Worthey (1994). We use the calibration of Worthey (1994) for the $R-K$ color assuming age 17 Gyr in order to measure metallicities for the clusters of our sample. We note that the available data for Galactic globular clusters (the $V-K$ data of Aaronson $et$ $al.$ 1978 in combination with the $V-R$ colors listed by Harris 1995) are consistent with this calibration.

The final abundances for the 19 clusters with $K \leq 17$, which have accurate enough colors, are also listed in Table 1. The clusters redder than $R-K = 3.5$ will be discussed in the next section.

Jablonka $et$ $al.$ (1995) confirm the metallicity scale of Harris $et$ $al.$ (1992) up to near solar values. Our metal abundances for the five clusters in common with Harris $et$ $al.$ (1992) also agree –within the combined errors– with their determinations ($\Delta[Fe/H] = 0.1 \pm 0.2$, with $rms = 0.5$ dex). However, on average, the clusters in the inner 3 kpc of NGC 5128 are more metal–rich than the outer clusters studied by Harris $et$ $al.$ (1992). The majority of the clusters in our sample have $[Fe/H] = -0.3 \pm 0.4$ dex.

Since our metallicities are in the same system as those of Harris $et$ $al.$ (1992), we combine both datasets in order to determine the existence of a metallicity gradient in NGC 5128. The result of this is shown in Figure 6, where we plot the dependence of metallicity with radial distance. Based on this Figure, we establish the presence of a metallicity gradient in the globular cluster system of this galaxy.

Figure 7 shows the metallicity distributions for the inner ($R \leq 5$ kpc), and outer

($R \geq 5$ kpc) clusters. This Figure shows that the metal–rich clusters are more concentrated than the metal–poor clusters. The different concentrations of these two populations, which also have different kinematics (Hui et al. 1995), is responsible for the observed metallicity gradient. A similar result was recently obtained by Geisler et al. (1996) for the globular cluster system of the elliptical galaxy NGC 4472.

It is worth comparing the clusters in the inner region of NGC 5128 with the well studied Pop. II clusters in the Galaxy, and M31. The location of Milky Way and M31 globular clusters (Frogel *et al.* 1980) is indicated in Figure 4. We note that independently of the metallicity calibration, the clusters of our sample have colors similar to Galactic bulge globular clusters, that are redder in $R - K$ and $J - K$ than typical halo globular clusters of the Milky Way.

Old and metal–rich globular clusters are found to be associated with bulges of galaxies in the Local Group, as discussed by Minniti (1995) for the Milky Way, and by Jablonka *et al.* (1992) and Ashman & Bird (1993) for M31. Red globular clusters like those discussed here are also observed in large numbers around elliptical galaxies (e.g. Zepf et al. 1995, Harris 1996, Kissler-Patig et al. 1996). From the presence of these clusters in the inner regions of NGC 5128, we conclude that at least one of the galaxies that may have collided was a bulge–dominated galaxy (*i.e.* an early type spiral or an elliptical galaxy). This view is supported by the fact that the stellar population in NGC 5128 is similar to that of a typical giant elliptical galaxy (van den Bergh 1976, Graham 1979).

Young clusters ($age \leq 8$ Gyr) have been detected near the edge of the dust lanes (van den Bergh 1976, Dufour *et al.* 1979, Graham 1979, Nicholson *et al.* 1992, Schreier *et al.* 1996). However, there is no evidence in our data for very young globular clusters *away from the dust lanes*.

## 5. The Intermediate Age Clusters

It is also worth comparing the clusters in the inner regions of NGC 5128 with those of the Magellanic Clouds. Both the Large and Small Magellanic Clouds have a rich system of intermediate–age clusters (see Olszewski 1995 for a recent review).

The presence of carbon–type stars in the asymptotic giant branch determines that intermediate–age clusters are bright in the IR (Mould & Aaronson 1980, Persson *et al.* 1983). These "IR–enhanced" clusters have redder optical and near–IR colors, and fainter optical magnitudes than typical Pop. II globular clusters. These clusters are classified as SWB type IV–VI (Searle *et al.* 1980), having ages of $t \sim 1 - 4$ $Gyr$ (Aaronson & Mould 1982).

In Figures 3 to 5 there is a group of $5 - 6$ clusters with colors consistent with SWB type IV–VI clusters. Note that none of the clusters in NGC 5128 studied by Frogel (1984) had carbon stars. However, his clusters are much brighter ($M_K \approx -13$) than typical SWB IV–VI clusters in the LMC ($M_K \approx -10$); whereas the reddest clusters in Figure 3 have magnitudes consistent with intermediate–age clusters like NGC 2209 in the LMC.

Being fainter than typical globular clusters, these intermediate–age objects have large photometric errors. However, neither the photometric errors nor the presence of differential reddening seems to be able to account for the location of these clusters in all the color–color and color–magnitude di-agrams (Figures 3 to 5). The direction of the reddening vector is indicated in these diagrams. For example, a surprisingly high differential reddening might move the clusters in the right direction in the optical–IR color–color diagram (Figure 4), but they would still be too red in H–K (Figure 5). Frogel, Mould & Blanco (1990) demonstrated that intermediate age globular clusters such as NGC 2209 are very red, particularly in H–K.

If these clusters were formed during the possible merger, then this episode must have occurred a few Gyrs ago, according to the age of SWB IV–VI clusters in the Magellanic Clouds. Limits to the age of the merger event are set by the presence of shells (Malin *et al.* 1983). These and other pieces of evidence favor a recent merger, having occurred less than a few Gyrs ago (Graham 1979, Schiminovich *et al.* 1995).

Alternatively, we may be looking at the cluster members of one of the colliding galaxies, in which case we would suggest that it was a late–type spiral (LMC or M33 type). This hypothesis is supported by the vast amounts of dust and gas present in the inner regions of NGC 5128.

## 6. Conclusions

We have identified 32 globular clusters in the inner 3 kpc of NGC 5128, the nearest known large galaxy being the possible product of a merger. The clusters have been selected on the basis of their structural parameters (observed core diameters and ellipticities), as measured from deep WFPC HST images. These data have been complemented with deep IR photometry obtained at the ESO/MPI 2.2–m telescope.

Most of the cluster candidates have magnitudes and colors similar to those of old globular clusters in the Milky Way and M31. Based on the optical and IR colors, the vast majority of these clusters is metal–rich ($-0.6 \leq [Fe/H] \leq +0.1$), significantly more metal–rich than the globular clusters in the outer regions. We therefore conclude that there is a metallicity gradient in the globular cluster system of NGC 5128.

A significant fraction of the candidate clusters have magnitudes and colors consistent with them being intermediate–age clusters, similar to SWB type IV–VI clusters that are found in the LMC and SMC. The colors of these clusters must be dominated by bright carbon stars, making the detection unambiguous. This represents the first detection of such objects beyond the Local Group, even though spectroscopic confirmation is desirable. Optical spectra should reveal the signature of carbon stars, namely strong Swan bands of $C_2$.

Finally, the finding of different cluster populations deep in the potential well of NGC 5128 offers the exciting possibility of constraining the formation of this galaxy. For example, from the presence of both metal–rich old globular clusters and intermediate–age clusters, we speculate that NGC 5128 was formed by the merging of a bulge dominated galaxy (elliptical or early–type spiral) with a late–type disk galaxy (like the LMC or M33), as originally proposed by Baade & Minkowski (1954), and Malin *et al.* (1983).

This work is dedicated to the memory of J. L. Sérsic. We thank R. Sharples for sharing with us his unpublished data on the globular clusters in the inner regions of NGC 5128, D. Geisler for advice regarding the transformation of the Washington photometry, and E. Olszewski for comments and suggestions about the clusters in the Magellanic Clouds. We also thank the referee S. Zepf for useful suggestions. M. V. Alonso acknowledges support from the ESO Visitor's Program. This work was performed in part under the auspices of the U. S. Department of Energy by Lawrence Livermore National Laboratory under Contract W-7405-Eng-48.

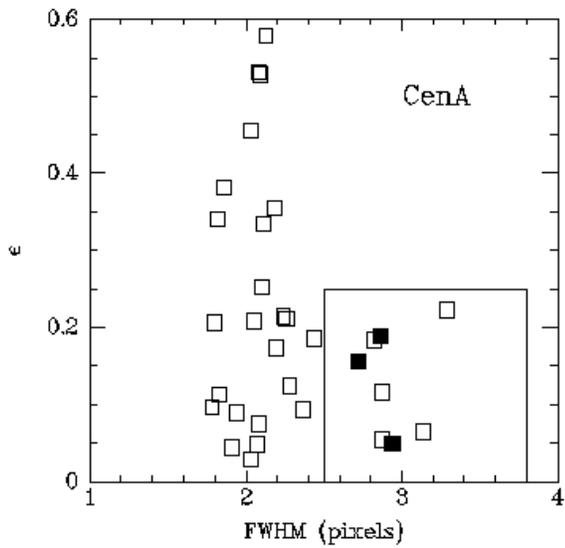 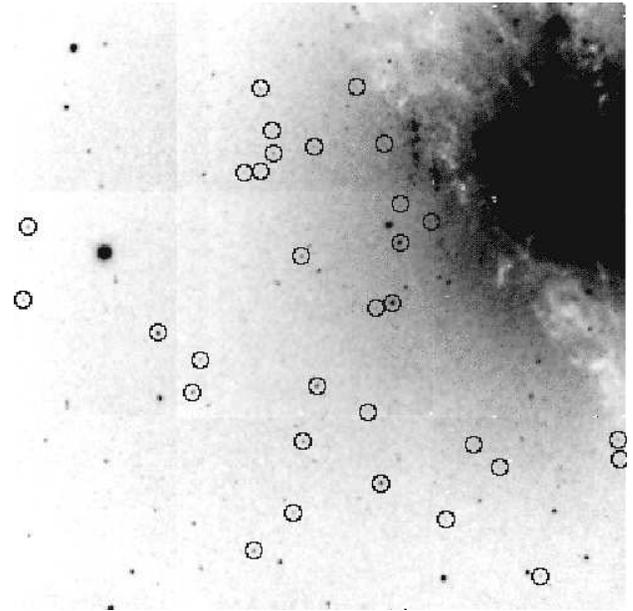

Fig. 1.— Ellipticity $\epsilon$ vs. FWHM of all the sources identified in the PC4 chip down to $R = 21$ *mag*. The loci of the globular clusters G206, G268 and G359 (Harris *et al.* 1992) are indicated (filled squares). The region used to select potential globular clusters in all the chips of WFPC is enclosed with the solid line.

Fig. 2.— $H$ mosaic of the central region of NGC 5128 observed with IRAC2B at the ESO/MPI 2.2–m telescope. This mosaic covers $\sim 3 \times 3$ kpc at the distance of NGC 5128. The location of the WFPC frames is shown in red. The globular cluster candidates are indicated with blue circles. West is up and North to the left.

TABLE 1
PHOTOMETRY OF NGC 5128 CLUSTERS

| ID | H92 | $x$ | $y$ | $r$ | $K$ | $\sigma_K$ | $J-K$ | $H-K$ | $R$ | $R-K$ | $[Fe/H]$ |
|---|---|---|---|---|---|---|---|---|---|---|---|
| 1  |      | 326.3 | 20.6  | 2.1 | 16.78 | 0.35 | 0.21  | 0.00  | 19.43 | 2.65 | −0.40 |
| 2  | G268 | 150.0 | 35.7  | 2.6 | 15.75 | 0.12 | 0.87  | 0.17  | 18.53 | 2.78 | −0.27 |
| 3  |      | 242.0 | 42.0  | 2.1 | ——    | ——   | ——    | ——    | 20.90 | ——   | ——    |
| 4  |      | 268.0 | 55.0  | 2.0 | 16.32 | 0.22 | 1.72  | 0.76  | 20.46 | 4.14 | ——    |
| 5  |      | 174.2 | 59.6  | 2.3 | 16.14 | 0.16 | 1.64  | 0.50  | 20.12 | 3.98 | ——    |
| 6  | G242 | 228.1 | 77.3  | 2.0 | 15.21 | 0.08 | 0.97  | 0.09  | 18.00 | 2.79 | −0.26 |
| 7  |      | 301.0 | 87.0  | 1.6 | 17.56 | 0.59 | 0.72  | −0.46 | 20.05 | 2.49 | −0.56 |
| 8  |      | 180.6 | 102.9 | 2.1 | 16.06 | 0.16 | 0.89  | 0.20  | 18.91 | 2.85 | −0.22 |
| 9  |      | 218.5 | 122.0 | 1.7 | 16.57 | 0.20 | 1.80  | 0.81  | 20.51 | 3.94 | ——    |
| 10 | G206 | 112.1 | 133.2 | 2.4 | 15.31 | 0.10 | 1.47  | 0.73  | 18.55 | 3.24 | 0.04  |
| 11 |      | 188.8 | 137.1 | 1.8 | 15.53 | 0.09 | 1.08  | 0.27  | 18.50 | 2.97 | −0.13 |
| 12 |      | 117.0 | 153.0 | 2.3 | 15.88 | 0.14 | 1.55  | 0.79  | 19.75 | 3.87 | ——    |
| 13 | G359 | 91.2  | 169.8 | 2.4 | 15.21 | 0.08 | 1.14  | 0.25  | 18.24 | 3.03 | −0.09 |
| 14 |      | 225.1 | 184.9 | 1.4 | 17.19 | 0.37 | −0.01 | −0.45 | 19.75 | 2.56 | −0.47 |
| 15 |      | 235.1 | 188.2 | 1.3 | 14.58 | 0.05 | 0.91  | 0.35  | 17.09 | 2.51 | −0.54 |
| 16 | G292 | 8.3   | 189.9 | 3.0 | 16.45 | 0.21 | 0.73  | 0.19  | 19.34 | 2.89 | −0.19 |
| 17 |      | 179.3 | 216.8 | 1.6 | 16.66 | 0.24 | 0.34  | −0.39 | 18.53 | 1.87 | −1.60 |
| 18 |      | 239.1 | 225.0 | 1.1 | 14.96 | 0.08 | 0.93  | 0.17  | 17.39 | 2.43 | −0.66 |
| 19 | G169 | 11.1  | 234.7 | 2.9 | 16.07 | 0.14 | 0.68  | 0.05  | 18.97 | 2.90 | −0.18 |
| 20 |      | 257.9 | 238.0 | 0.9 | 16.19 | 0.25 | 0.68  | 0.43  | 20.05 | 3.86 | ——    |
| 21 |      | 245.5 | 252.0 | 1.0 | 17.99 | 1.29 | 0.53  | ——    | 19.38 | 1.39 | −2.0: |
| 22 |      | 144.2 | 268.2 | 1.8 | 17.93 | 0.74 | 0.84  | ——    | 20.05 | 2.12 | −1.28 |
| 23 |      | 154.0 | 269.0 | 1.7 | ——    | ——   | ——    | ——    | 20.51 | ——   | ——    |
| 24 |      | 162.0 | 280.0 | 1.7 | 16.56 | 0.26 | 1.14  | 0.12  | 19.62 | 3.06 | −0.06 |
| 25 |      | 186.8 | 284.1 | 1.5 | ——    | ——   | ——    | ——    | 19.17 | ——   | ——    |
| 26 |      | 231.0 | 285.0 | 1.1 | 17.03 | 0.64 | 0.67  | −0.12 | 20.08 | 3.05 | −0.06 |
| 27 |      | 161.0 | 294.0 | 1.7 | ——    | ——   | ——    | ——    | 20.81 | ——   | ——    |
| 28 |      | 153.6 | 319.7 | 1.8 | 16.77 | 0.28 | 0.48  | −0.03 | 19.34 | 2.57 | −0.46 |
| 29 |      | 291.0 | 97.0  | 1.6 | ——    | ——   | ——    | ——    | 21.91 | ——   | ——    |
| 30 |      | 368.0 | 101.0 | 1.4 | ——    | ——   | ——    | ——    | 20.74 | ——   | ——    |
| 31 |      | 368.0 | 107.5 | 1.4 | 17.46 | 0.79 | 1.46  | ——    | 21.83 | 4.37 | ——    |
| 32 |      | 212.0 | 322.0 | 1.3 | 16.98 | 0.37 | 0.35  | −0.30 | 20.71 | 3.73 | ——    |

REFERENCES.— H92 = Harris *et al.* 1992.
The $x$ and $y$ coordinates are in pixels, with scale 0.49 $arcsec/pix$.
The projected radial distance from the nucleus, $r$, is in arcminutes.

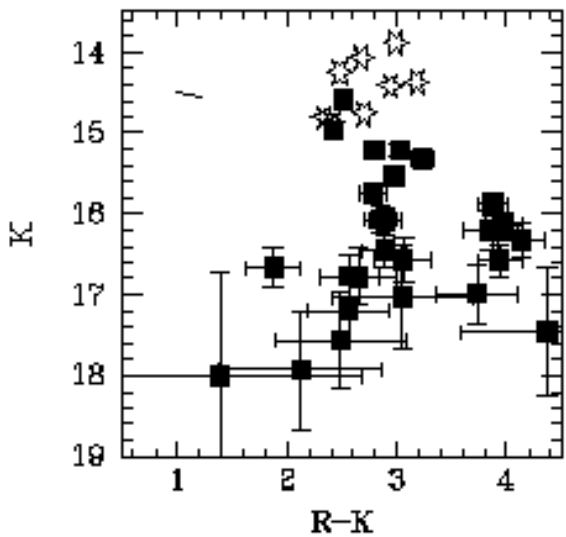

Fig. 3.— $K$ vs. $R - K$ color–magnitude diagram for the globular cluster candidates from this work (squares), in combination with the globular clusters studied by Frogel (1984) (stars). The solid line is the reddening vector corresponding to $E(B - V) = 0.11$.

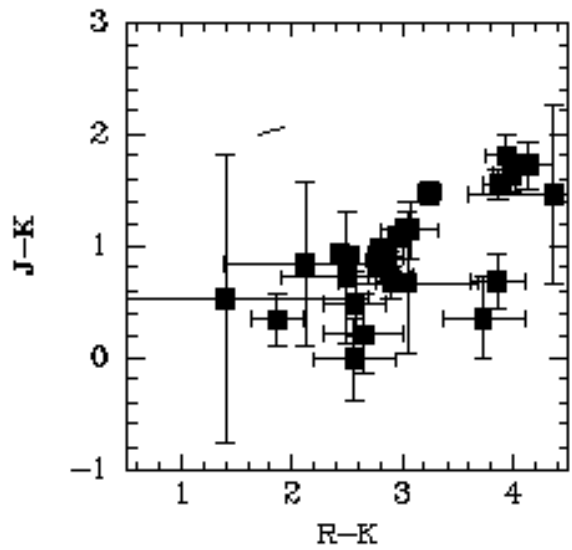

Fig. 4.— Optical–IR color–color diagram for our globular cluster candidates. The location of intermediate–age clusters in the LMC is indicated. One–$\sigma$ errorbars are shown. The solid line is the reddening vector corresponding to $E(B - V) = 0.11$.

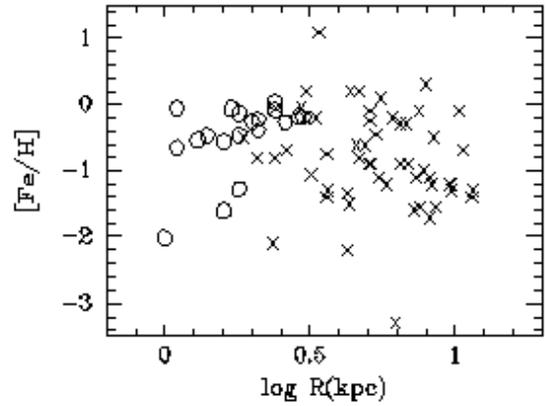

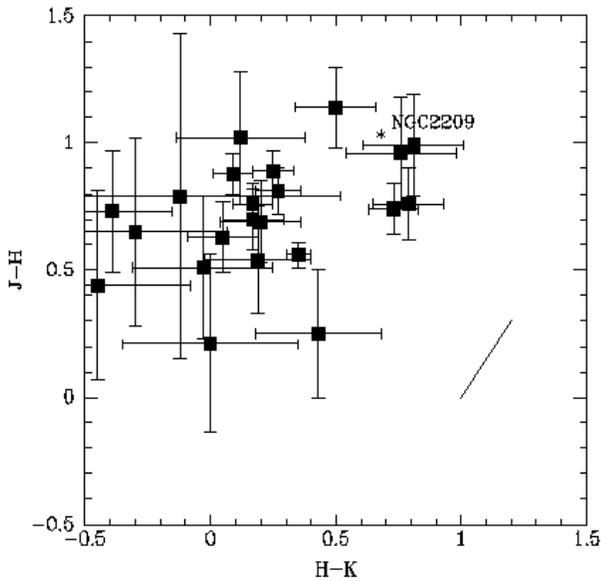

Fig. 6.— Radial dependence of metallicity for globular clusters in the inner regions of NGC 5128 from Table 1 (open circles), and from Harris et al. (1992, crosses). This Figure shows that the metal–rich clusters are more concentrated than the metal–poor ones.

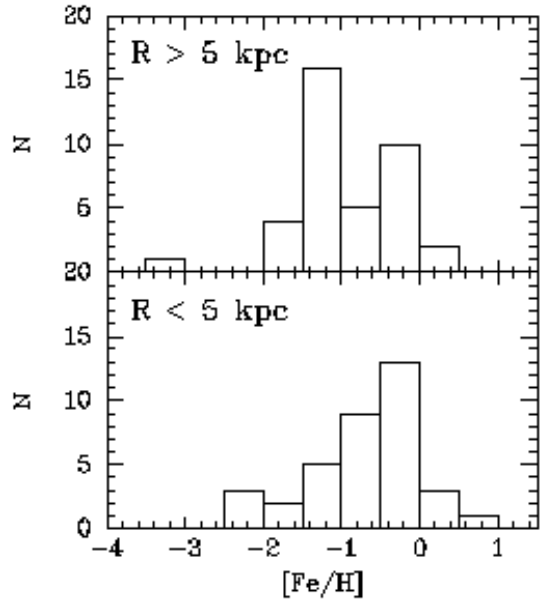

Fig. 5.— IR color–color diagram for the globular cluster candidates. The location of the intermediate–age cluster NGC 2209 in the LMC is indicated. The solid line is the reddening vector corresponding to $E(B-V) = 1.1$, 10 times the reddening value adopted here. The one–$\sigma$ errorbars are shown.

Fig. 7.— Metallicity distribution for the outer and inner clusters in NGC 5128 from Harris et al. (1992) and this work. Note that the size of the bins is larger than the estimated errors in the metallicities.